# New formalism for selfconsistent parameters optimization of highly efficient solar cells

A.V. Sachenko[1], V.P. Kostylyov[1], M.R. Kulish[1], I.O. Sokolovskyi[1], A. Chkrebtii[2]

[1] V. Lashkaryov Institute of Semiconductor Physics, NAS of Ukraine

41, prospect Nauky, 03028 Kyiv, Ukraine

[2] University of Ontario Institute of Technology, Oshawa, ON, Canada
E-mail: sach@isp.kiev.ua

Abstract

We analysed self-consistently photoconversion efficiency of direct-gap A3B5 semiconductors based solar cells and optimised their main physical characteristics. Using gallium arsenide (GaAs) as the example and new efficient optimization formalism, we demonstrated that commonly accepted light re-emission and re-absorption in solar cells (SC) in technologically produced GaAs (in particular, with solid- or liquid-phase epitaxy) are not the main factors responsible for high photoconversion efficiency. As we proved instead, the doping level of the base material and its doping type as well as Shockley-Read-Hall (SRH) and surface recombination velocities are much more important factors responsible for the photoconversion. We found that the maximum photoconversion efficiency (about 27% for AM1.5 conditions) in GaAs with typical parameters of recombination centers can be reached for p-type base doped at $2 \cdot 10^{17}$ cm$^{-3}$.

The open circuit voltage $V_{OC}$ formation features are analyzed. The optimization provides a significant increase in $V_{OC}$ and the limiting photoconversion efficiency close to 30%.

The approach of this research allows to predict the expected solar cells (for both direct-gap and indirect-band semiconductor) characteristics if material parameters are known. Obtained formalism allows to analyze and to optimize mass production both tandem solar cell (TSC) and one-junction SC parameters.

**Introduction**

During the last years, an approach to the modeling of the tandem solar cells (TSC) limit photoconversion efficiency that considers photon recycling (re-emission and reabsorption of photons) in structures with multiple absorption of light as the key point was developed (see, eg, [1,2]). A point of view that there is a strong correspondence between the internal quantum yield of luminescence $q_{pl}$ and photoconversion efficiency $\eta$ (for defined model of the light absorption in the structure) is presented in these researchs. The photoconversion efficiency in GaAs was calculated in papers [1,2] as a confirmation of this concept. This material is not chosen by chance. On the one hand, the GaAs is the key semiconductor when creating TSC based on direct-gap semiconductors $A^3B^5$ group. On the other hand, in double heterostructures AlGaAs - GaAs – AlGaAs the internal luminescence quantum yield is close to 100% under certain conditions as was shown experimentally [3]. Therefore, the influence of reabsorption and re-emission of light on the efficiency of solar cells based on GaAs announced very significant [1,2].

However, as will be shown below, idealized assumptions used in [1,2] are not being implemented in practice, and the implementation of typical semiconductor solar cell parameters are unsuitable for calculating and optimizing the parameters of a solar cell. Let us consider these assumptions in more detail. The basic assumption is the neglect of nonradiative recombination channels. The internal quantum yield of luminescence close to 100% is assumed. In fact, such value of $q_{pl}$ was obtained in laser structure with a high excitation level (~ $10^{17}$ cm$^{-3}$) [3]. In standard operating SC conditions the excitation is at several orders of magnitude smaller. It leads to a substantial decrease in the internal quantum efficiency of luminescence. To prove this, we obtained, solved and analysed in details the equation for the magnitude $q_{pl}$ which is valid for an arbitrary excitation level and takes into account the basic mechanisms of recombination.

Next approximation [1,2] is the assumption that the absorption coefficient $\alpha$ of light in high-purity gallium arsenide (doping level of the semiconductor $\leq 10^{14}$ cm$^{-3}$) can be used for the calculation of short-circuit current and open circuit voltage. In fact, as shown in this paper, even if implemented currently longest in GaAs Shockley -Read - Hall lifetimes (~ $10^{-6}$ s), to get large values of the open circuit voltage it is necessary to implement the doping level of the semiconductor of $\geq 10^{17}$ cm$^{-3}$. Thus, as shown in [5], the light absorption coefficient of doped GaAs is substantially different from absorption coefficient of the high-clean GaAs. As a result, the short-circuit current in the p-type semiconductor increases and in the n-type semiconductor decreases.

In the calculation of short-circuit current expressions for the internal quantum efficiency $q_{SC}$ for cases where diffusion length $L$ significantly exceeds SC thickness $d$ and of the surface recombination velocity $S$ is zero are used [1,2]. In this article much more general expression for the $q_{SC}$ is used for plane-parallel structures. It is valid in particular for arbitrary ratio between $L$ and $d$, for arbitrary $S$ and for any value of the light reflectance on the back surface $R_d$, varying from 0 to 1.

Ding et. al. [4] utilize more general approach then [1,2]. The calculation of GaAs photo-conversion efficiency [4] takes into account the recombination in the space charge region (SCR) in the recombination current calculation as well as the reabsorption and re-emission of photons. The saturation current $I_{0r}$ is introduced for the implementation of I-V curve non-ideality factor equal to 2. However, calculation [4] does not take into account Shockley -Read - Hall recombination current in the neutral part of base region. and there is no connection between the saturation current $I_{0r}$ in [4] is not a function of such physical parameters as Shockley -Read - Hall recombination time $\tau_{SR}$, the doping level of the semiconductor, the effective densities of states in the conduction and valence bands and the band gap. The model [4] does not allow to calculate the $I_{0r}$ value for a particular semiconductor and regarded structure parameters. It denies SC optimization by the recombination current minimization.

The extended and modified GaAs SCs efficiency analysis in this paper uses some results of [5]. In particular, this research is correct for the case of an arbitrary injection level, i.e. for arbitrary excess (generated by the light) electron-hole pairs concentration $\Delta p$. The luminescence internal quantum yield for the arbitrary $\Delta p$ is analyzed regarding main recombination mechanisms. The space charge region (SCR) recombination velocity and current corrected equations are determined and used to calculate open-circuit voltage. The radiative recombination coefficient $A$ is calculated using Roosbroeck and Shockley equations [6]. This work introduces selfconsistent analysis of factors influencing photoconversion efficiency in more general assumptions than [1, 2, 4].

In our work the equation for open-circuit voltage is written regarding the neutral part of the base region and SCR recombination currents. Saturation currents take into account radiative and non-radiative recombination and other semiconductor parameters so the specific value of this ratio can lead to both the increase and the decrease in this current. It is shown that this dependence in some cases is essential for the SC photovoltage calculation.

It is shown that the main SC parameters could not been chosen arbitrarily, because they are interdependent. For example, determining the doping level, it is necessary to consider the change of light absorption coefficient and the radiative recombination coefficient. It is found that reabsorption and re-emission of light are not major factors for the GaAs SCs efficiency.

These results can be used to select the optimal parameters for SCs produced using both gallium arsenide and other semiconductor materials industrial technologies (including germanium and silicon solar cells) and to achieve maximum photoconversion efficiency.

# 1. Basic relations and the analysis of the influence of various factors on the GaAs photoconversion efficiency

Let us analyze GaAs photoconversion efficiency regarding the doping level and conductivity type, deep level carriers capture cross sections ratio, radiative and non-radiative recombination probabilities ratios for the SCR and the neutral part of the base region.

## 1.1. Influence of doping level and conductivity type

Let us analyze the behavior of the light absorption coefficient $\alpha$ of GaAs as a function of photon energy $E_{ph}$, depending on the conductivity type and the doping level of the semiconductor. These dependences $\alpha(E_{ph})$ are shown in Fig. 1 [5]. Figure 1 shows the increase of the fundamental absorption edge blurry with the increase of doping level in p-type GaAs (Urbach effect), i.e. absorption increase in the spectral region where the photon energy $E_{ph}$ is less than band gap Eg ($E_{ph} < E_g$). At the same time in n-type GaAs at typical for TSCs levels of doping there is a degeneration of GaAs semiconductor and Burstein-Moss effect domination. As a result the doping level increase shifts the absorbtion edge to the high-energy region. Thus the short-circuit current for a given thickness $d$ decreases.

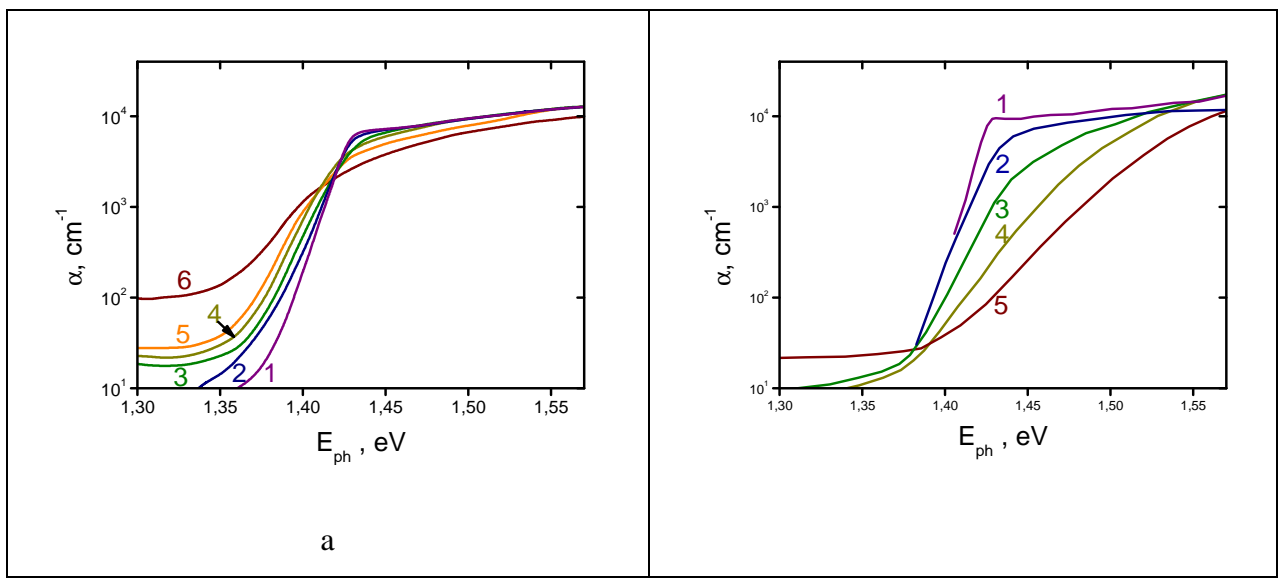

Figure 1. The doping level influence on the absorption edge of p-type (a) and n-type (b) GaAs [5]. P-type doping level, cm$^{-3}$: 1 – 1.6·10$^{16}$, 2 - 2.2·10$^{17}$, 3 - 4.9·10$^{17}$, 4 - 1.2·10$^{18}$, 5 - 2.4·10$^{18}$, 6 - 1.6·10$^{19}$; n-type doping level, cm$^{-3}$: 1 – 5·10$^{13}$, 2 - 5.9·10$^{17}$, 3 - 2·10$^{18}$, 4 -3.3·10$^{18}$, 5 - 6.7·10$^{18}$

Using the known $\alpha(E_{ph})$ dependence [5] and the formula [6]

$$A = \frac{(kT)^3}{\pi^2 c^2 \hbar^3 n_i^2} \int_0^\infty \frac{\varepsilon(u)\alpha(u)u^2 du}{e^u - 1}, \qquad (1)$$

we can calculate the radiative recombination coefficient A. Here c is the speed of light, $\hbar$ is Planck's constant, $n_i$ is the intrinsic concentration of semiconductor, $\varepsilon(u)$ is dielectric permittivity of the semiconductor, taking into account the dispersion of the photon energy, $u = E_{ph}/kT$ is the dimensionless photon energy.

The calculated values of A are shown in Figure 2. The figure shows that in the p-type semiconductors with doping level increase, the value of A increases from $7 \cdot 10^{-10}$ cm$^3$/s for $p_0 = 10^{17}$ cm$^{-3}$ to $1.5 \cdot 10^{-9}$ cm$^3$/s for $p_0 = 2 \cdot 10^{18}$ cm$^{-3}$. At the same time, in $n$-type semiconductors with doping level increase, the value of A decreases from $10^9$ cm$^3$/s for $n_0 = 10^{14}$ cm$^{-3}$ to $9 \cdot 10^{-11}$ cm$^3$/s for $n_0 = 2 \cdot 10^{18}$ cm$^{-3}$.

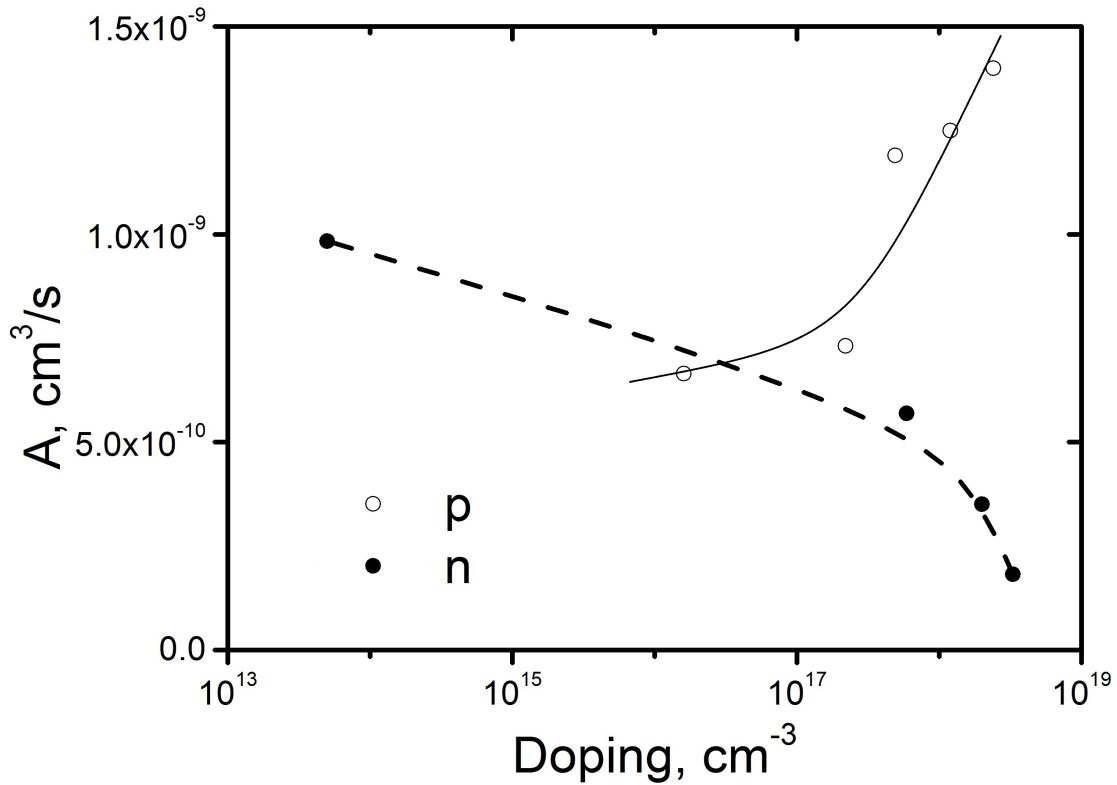

Figure 2. The radiative recombination coefficient $A$ versus the doping level in the n- and p-type GaAs.

The influence of doping level on the value of bulk lifetime in GaAs [8-10] is shown in Fig. 3. In this case

$$\tau_b = \left(\frac{1}{\tau_{SR}} + \frac{1}{\tau_r}\right)^{-1}, \qquad (2)$$

where $\tau_{SR}$ is the Shockley-Read-Hall, $\tau_r = (A(N_d)N_d)^{-1}$ is radiative lifetime and $N_d$ is the doping level of the base. Figure 3 shows that the bulk lifetime in p-type GaAs is lower than in $n$-type. Data of the Fig. 3 correlate with the data on the lifetimes for n-and p-type GaAs [9].

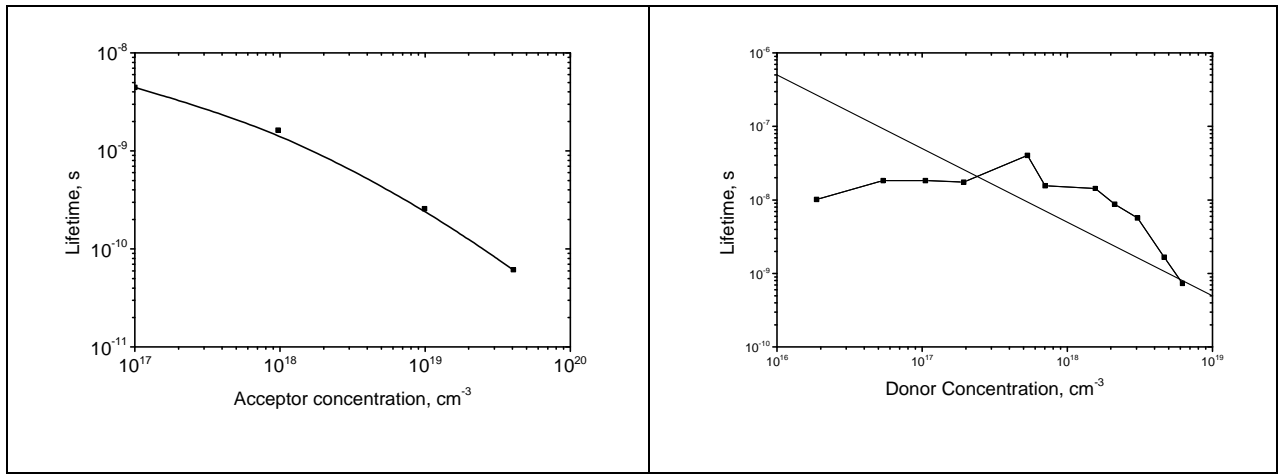

Figure. 3. Bulk lifetime in p- and n-type GaAs versus doping level [5]. Points are experimental values. Line in the figure for the n-type corresponds to the constant radiative recombination lifetime value.

The contributions of $\tau_r$ and $\tau_{SR}$ can be divided using the known values of $A(N_d)$ and the relation (2). This way the values of $\tau_{SR}$ for $n$- and p-type GaAs with different doping levels can be achieved. Typical $\tau_{SR}$ value is equal to $2 \cdot 10^{-8}$ s for $n$-type GaAs in doping levels range from $1 \cdot 10^{16}$ cm$^{-3}$ to $5 \cdot 10^{18}$ cm$^{-3}$. For the doping level $n_0$ of $10^{16}$ cm$^{-3}$ radiative lifetime $\tau_r$ is equal to $10^{-7}$ s, while for $n_0 = 2 \cdot 10^{18}$ cm$^{-3}$ $\tau_r = 2.5 \cdot 10^{-8}$ s. In the first case $\tau_{SR}$ is less than $\tau_r$, while in the second case the times $\tau_{SR}$ and $\tau_r$ are of the same order, i.e. radiative recombination does not dominate. For this case photon recycling will be weakened even in highly reflective structures. Moreover, regarding to the photon recycling [4], the value of radiative recombination coefficient decreases. In this case, an effective radiative recombination coefficient $A_{eff} = A(1 - \gamma_r)$ (here $\gamma_r$ is the photon recycling factor) should be used in the open-circuit volt-age $V_{OC}$ calculation. This effect leads to the increase in radiative lifetime and to the raise of nonradiative recombination.

Typical $\tau_{SR}$ values in the p-type semiconductors at low excitation levels ($\Delta p \ll p_0$) are about $5 \cdot 10^{-9}$ s [8,9]. At sufficiently high levels of excitation $\tau_{SR}$ is about $2.5 \cdot 10^{-7}$ s. A simple calculation allows to determine using these values the ratio of the electrons and holes capture cross sections on recombination level, which in this case is equal to fifty.

Thus, as follows from the analysis of known data about Shockley-Read-Hall lifetimes $\tau_{SR}$, for typical $\tau_{SR}$ values in n- and p-type GaAs, n-type GaAs with greater $\tau_{SR}$ suites as the base material better. This doping type is also proposed in Kayes patent [10].

## 2.2. Internal quantum efficiency of the short circuit current

More general expression for the quantum yield of the short-circuit current $q_{sc}(E_{ph})$ [11] can be written as

$$q_{sc} = q_{sp} + q_{sn}, \qquad (3)$$

$$q_{sp} = \frac{\alpha L_p}{(\alpha L_p)^2 - 1} \cdot \frac{\alpha L_p + S_0 \frac{\tau_p}{L_p}\left(1 - e^{-\alpha d_p}\right)\cosh\left(\frac{d_p}{L_p}\right) - e^{-\alpha d_p}\sinh\left(\frac{d_p}{L_p}\right) - \alpha L_p e^{-\alpha d_p}}{S_0 \frac{\tau_p}{L_p}\sinh\left(\frac{d_p}{L_p}\right) + \cosh\left(\frac{d_p}{L_p}\right)}, \qquad (4)$$

$$q_{sn} = \frac{\alpha L e^{-\alpha d_p}}{1 - (\alpha L)^2} \cdot \left\{ \frac{\left[S_d \cosh\left(\frac{d}{L}\right) + \frac{D}{L}\sinh\left(\frac{d}{L}\right)\right]\left(1 + R_d e^{-2\alpha d}\right) + (\alpha D(1 - R_d) - S_d(1 + R_d))e^{-\alpha d} -}{S_d \sinh\left(\frac{d}{L}\right) + \frac{D}{L}\cosh\left(\frac{d}{L}\right)} \right.$$

$$\left. \frac{- \alpha L S_d\left[\sinh\left(\frac{d}{L}\right) + \frac{D}{L}\cosh\left(\frac{d}{L}\right)\right]\left(1 - R_d e^{-2\alpha d}\right)}{S_d \sinh\left(\frac{d}{L}\right) + \frac{D}{L}\cosh\left(\frac{d}{L}\right)} \right\}. \qquad (5)$$

Here $\alpha$ is the light absorption coefficient, $L_p$ is diffusion length in the emitter, $d_p$ is the thickness of the emitter, $\tau_p$ is the bulk lifetime in the emitter, $R_d$ is light reflection coefficient of the back surface, $L = (D\tau_b)^{1/2}$ is the diffusion length in the base, $D$ is the diffusion coefficient in the base, $S_0$ is the effective surface recombination velocity at the surface of the emitter, $S_d$ is the effective surface recombination velocity at the back surface.

When the $L \gg d$ criterion fulfils and when $S_0$ and $S_0$ are small (compared to the recombination rate in the base volume), expression for $q_s(E_g, E_{ph})$ reduces to $1-\exp(-\alpha d)$ expression for $R_d \to 0$ (poorly reflecting structures of type A (Fig. 4)), and to expression $1-\exp(-2\alpha d)$ for $R_d \to 1$ (highly reflective structures of type B (Fig. 4) [4]). However, for sufficiently low recombination times when $L \leq d$, as well as for large values of $S_0$ and $S_d$ the short-circuit current quantum yield can be significantly reduced relative to the values above.

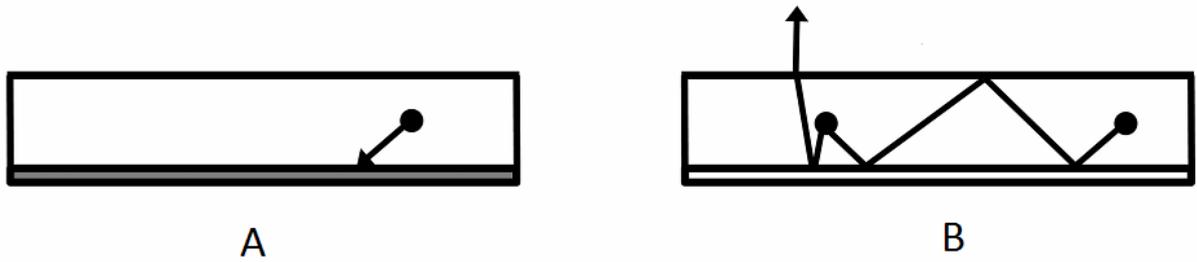

Fig. 4 Schematic diagrams of two types of planar solar cell structures, labeled Structures A and B. Structure A is a semiconductor slab with an absorbing substrate, Structure B is a semiconductor slab with a reflecting substrate. Source: [4]

## 2.3. The open circuit voltage

We obtained more definite, than the commonly used, expression for the open circuit voltage $V_{OC}$ of solar cell with the $p-n$ junction. We suppose that in this expression the recombination current consists of two components: the neutral region recombination current and the base region SCR recombination current. Here we assume that Shockley-Read recombination goes through the discrete level close enough to the middle of the bandgap. Expressions for the current recombination in the SCR ([12] p. 104) can not be used directly for the illuminated SC. Furthermore, they are obtained in assumption of the equality of recombination centers capture cross sections of electrons and holes. We determine the contribution of SCR recombination in the open circuit voltage $V_{OC}$ and in the internal quantum yield of luminescence $q_{pl}$ more correctly for a realistic case when cross sections are not equal. At first we obtain the value of the recombination velocity in the SCR $V_{SC}$ for the illuminated SC when recombination in the SCR goes through a deep recombination level. We assume the concentration of recombination centers $N_r$, the capture cross section of electrons $\sigma_n$ and holes $\sigma_p$, energy placement is taken relative to the midgap. For n-type semiconductor under illumination, we can write

$$V_{SC}(\Delta p) = \frac{L_D}{\tau_{SR}} \int_{y_{pn}}^{0} \frac{n_0}{\left[(n_0 e^y + n_i e^{\varepsilon_r}) + b(p_0 + \Delta p)e^{-y} + n_i e^{-\varepsilon_r}\right]\sqrt{-y + (e^y - 1) + \frac{(p_0 + \Delta p)}{n_0} e^{-y}}} dy$$

. (6)

Here $L_D = \left(\varepsilon_0 \varepsilon_s kT / 2q^2 n_0\right)^{1/2}$ is the Debye screening length, $\varepsilon_0$ is the dielectric constant of vacuum, $\varepsilon_s$ is the relative permittivity of the semiconductor, $k$ is the Boltzmann constant, $T$ is a temperature, $q$ is the elementary charge, $\Delta p$ is the minority carriers excess concentration in the base, $\tau_{SR} = \left(C_p N_r\right)^{-1}$ is the Shockley-Read-Hall lifetime, $C_p = V_T \sigma_p$, $C_n = V_T \sigma_n$, $b = \sigma_p / \sigma_n$ is carriers cross sections ratio, $V_T$ is carriers average thermal velocity, $y_{pn}$ is the dimensionless potential at the $p-n$-junction interface, $y$ is the actual dimensionless (normalized to $kT/q$) electrostatic potential,

$$n_i(T) = \sqrt{N_c N_v}\left(\frac{T}{300}\right)^{3/2} \exp\left(-\frac{E_g}{2kT}\right) \quad (7)$$

is the intrinsic carriers concentration, $N_c$ and $N_v$ are effective densities of states in the conduction band and valence band for T =300 K; $E_g$ is the bandgap and $\varepsilon_r = E_r / kT$.

The same deep recombination center responsibility for recombination in SCR and for Shockley-Read recombination in the neutral base region of the SC is assumed in (6) for simplicity.

$V_{SC}(\Delta p)$ calculation shows that for $n_0 \approx 10^7$ cm$^{-3}$ and $\Delta p$ ranged from $10^{12}$ cm$^{-3}$ to $10^{14}$ cm$^{-3}$, corresponding to the SC working conditions for different Shockley-Read-Hall recombination times, the recombination rate does not depend on the $E_r$ value up to $\pm 0.2$ eV. Thus, when the $E_r$ modulo value is less than 0.2 eV, the terms proportional to $n_i$ in the denominator of expression (6) integrand can be neglected. It means that under these conditions the expression (6) for the SCR recombination velocity $V_{SC}(\Delta p)$, written for deep recombination centers, located in the middle of the gap, is correct for the high enough separation (within 0.2 eV) of the recombination level from the middle of the bandgap.

Using the expression (6) for $E_r = 0$ case, when subintegral function has a maximum and $y_m = \ln(n_0 / b(p_0 + \Delta p))/2$ SCR recombination velocity $V_{SC}$ can be written approximately as

$$V_{SC} = \frac{\kappa L_D}{\tau_{SR}} \frac{\left(\frac{n_0/b}{(p_0+\Delta p)}\right)^{1/2}}{\sqrt{\frac{1}{2}\ln\left(\frac{n_0/b}{(p_0+\Delta p)}\right)}}. \qquad (8)$$

Here, $\kappa$ is the constant of the order of 1. Calculations (1), (3) shows $\kappa$ increase from 1.9 to 2 with $\Delta p$ increase from $10^{10}$ cm$^{-3}$ to $10^{14}$ cm$^{-3}$. In this case the calculation results by the exact formula and approximate evaluation coincide with less then 5% error.

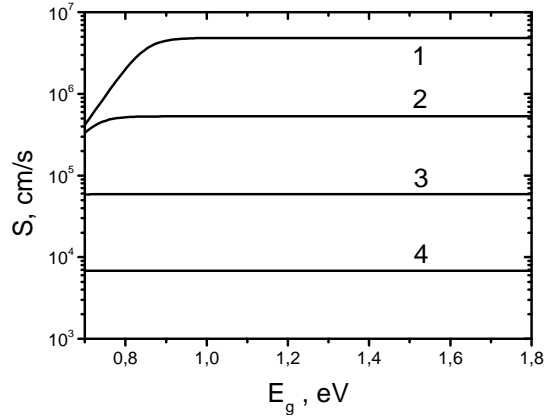

Figure. 5. The recombination velocity in the SCR versus the bandgap for following parameters: $\tau_{SR} = 10^{-7}$ s, $n_0 = 10^{17}$ cm$^{-3}$, $T = 300$ K, $b = 10^{-2}$, $\Delta p$, cm$^{-3}$: 1 – $10^7$; 2 – $10^9$; 3 – $10^{11}$; 4 – $10^{13}$.

Fig. 5 shows the calculated $V_{SC}(E_g)$ dependences constructed by formula (8). The parameter is the concentration $\Delta p$ of minority holes in the base. One can see from the figure that starting from $10^{10}$ cm$^{-3}$ $V_{SC}$ is not dependent on $E_g$. $V_{SC}$ value decreases with $\Delta p$ increase.

When illumination makes $\Delta p \gg p_0$ the relationship between $V_{SC}$ and recombination saturation current density $J_{rs}$ is described by the relation [13]

$$V_{SC} = \frac{J_{rs}}{q}\left(\frac{\Delta p}{p_0}\right)^{1/2} \Delta p^{-1}, \qquad (9)$$

$$J_{rs} \approx \frac{q \kappa L_D b^{-1/2} n_i}{\tau_{SR}\sqrt{\frac{1}{2b}\ln\left(\frac{qn_0(d/\tau_b + S_s)}{J_g}\right)}}, \qquad (10)$$

where $S_s = S_0 + S_d$.

Generation-recombination balance equation ignoring the SCR recombination

$$J_{SC} \approx q\left(\frac{d}{\tau_b} + S_s\right)\Delta p \qquad (11)$$

was used in (10). The logarithm parameter (10) is much greater then 1, so the first approximateon can be used. Here $J_{SC}$ is the short-circuit photocurrent density. The equation (11) is correct when the diffusion length $L$ is greater then the base thickness $d$.

Taking into account SCR recombination the open circuit voltage $V_{OC}$ can be found as

$$J_g = q\left(\frac{d}{\tau_b} + S_s\right)\frac{n_i^2}{n_0}\left[\exp\left(\frac{qV_{oc}}{kT}\right) - 1\right] + q\frac{\kappa L_D b^{-1/2} n_i}{\tau_{SR}\sqrt{\frac{1}{2}\ln\left(\frac{1}{b}\frac{qn_0(d/\tau_b + S_s)}{J_g}\right)}}\left[\exp\left(\frac{qV_{oc}}{2kT}\right) - 1\right] \quad (12a)$$

when the second term is not much greater then the second. If the second term is much more greater then the first

$$J_g = q\left(\frac{d}{\tau_b} + S_s\right)\frac{n_i^2}{n_0}\left[\exp\left(\frac{qV_{oc}}{kT}\right) - 1\right] + q\frac{\kappa L_D b^{-1/2} n_i}{\tau_{SR}\sqrt{\ln\left(\frac{\tau_{SR}}{q\kappa}\frac{J_g}{L_D n_i \exp(\frac{qV_{oc}}{2kT})}\right)}}\left[\exp\left(\frac{qV_{oc}}{2kT}\right) - 1\right] \quad (12b)$$

should be used to find the open-circuit voltage $V_{OC}$.

### 2.4. The photoluminescence internal quantum yield

Let us write the expression for the photoluminescence internal quantum yield regarding the radiative recombination, the bulk Shockley - Read - Hall recombination, the Auger recombination, the surface recombination and the SCR recombination for arbitrary excitation level $\Delta p$. We consider the p-type semiconductor in the calculation. We take into account both the Shockley - Read - Hall time and the surface recombination velocity dependence on the excitation value for $b \neq 1$. Let us consider such nonradiative recombination parameters dependences on the equilibrium majority carriers concentration $p_0$, minority carrier concentration $n_0$ and excitation level $\Delta p$ as

$$\tau_{SR} = \tau_0\left(1 + b_b^{-1}\frac{n_0 + \Delta p}{p_0 + \Delta p}\right)^{-1}; \quad S = S_0\left(1 + b_s^{-1}\frac{n_0 + \Delta p}{p_0 + \Delta p}\right)^{-1};$$

$$S_{sc} = S_{sc0}\left(\frac{b_b n_0}{\Delta p}\right)^{1/2}, \quad R_{Auger} = C(p_0 + \Delta p)^2. \quad (13)$$

Here $b_b$ and $b_s$ are capture cross sections relations for the bulk and the surface recombination levels respectively, $R_{Auger}$ is the Auger recombination rate, $C$ is the Auger recombination coefficient.

This way $\tau_{SR}$ and $S$ dependences on $\Delta p$ and other parameters of (13) are typical for the Shockley - Read - Hall recombination. $S_{SC}$ dependence on $\Delta p$ and other parameters is determined by (8).

Then in the case when $L > d$ (as required for efficient photoconversion) the expression for the luminescence internal quantum efficiency can be written as

$$q_{pl} = \frac{\frac{d}{\tau_r}}{d\left(\tau_{SR}^{-1} + \tau_r^{-1}\right) + S + S_{sc} + R_{Auger}}. \quad (14)$$

Fig. 6 shows the calculated dependence of the luminescence internal quantum yield versus excitation level using the following parameters: $p_0 = 3 \cdot 10^{17}$ cm$^{-3}$, $S_0 = 1 \cdot 10^2$ cm/s, $b_p = b_s = 0.02$ [9], $d = 1$ μm, $C = 7 \cdot 10^{-30}$ cm$^6$/s [3], $S_{sc0}$ is calculated using (7). The following $\tau_0$ values were used for 1-4 curves respectively: $5 \cdot 10^{-9}$, $2 \cdot 10^{-8}$, $5 \cdot 10^{-8}$ and $10^{-6}$ s. Figure 6 shows the luminescence internal quantum yield increase with the increase in $\tau_0$ value. When $\tau_0 = 10^{-6}$ s, $q_{pl}$ value at the maximum is equal to 99.4%, which is very close to the value obtained in [3].

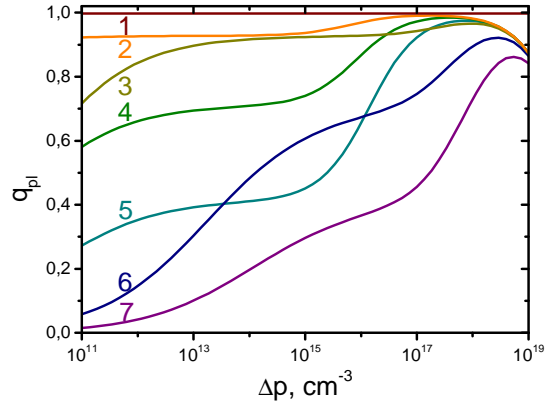

Figure 6. The luminescence internal quantum yield versus the injection level. Parameters used: $p_0 = 3 \cdot 10^{17}$ cm$^{-3}$, $d = 10^{-4}$ cm, $A = 5 \cdot 10^{-10}$ cm$^3$/c, $S = 10^3$ cm/s, 1 – $\tau_{SR} = 10$ s, $b = 0.02$; 2 – $\tau_{SR} = 10^{-6}$ s, $b = 0.02$; 3 – $\tau_{SR} = 10^{-6}$ s, $b = 50$; 4 – $\tau_{SR} = 2 \cdot 10^{-8}$ s, $b = 0.02$; 5 – $\tau_{SR} = 2 \cdot 10^{-8}$ s, $b = 50$; 6 – $\tau_{SR} = 5 \cdot 10^{-9}$ s, $b = 0.02$; 7 – $\tau_{SR} = 5 \cdot 10^{-9}$ s, $b = 50$.

However $\Delta p$ value in the SC under AM1.5 illumination is significantly lower. $\Delta p$ is determined by generation-recombination balance equation taking into account all recombination components:

$$J_{SC}/q = \left[d\left(\tau_{SR}^{-1} + \tau_r^{-1}\right) + S + S_{sc} + R_{Auger}\right]\Delta p. \quad (15)$$

Applied transcendent equations were solved numerically.

Fig. 7 shows $\Delta p$ dependence on the Shockley – Read – Hall lifetime for $A$ values of $5 \cdot 10^{-10}$ cm$^3$/s and $10^{-11}$ cm$^3$/s. It can be seen that the curves for different $A$ values coincide when $\tau_{SR} \leq 10^{-8}$ s. The reason for this coincidence is in that $\Delta p \propto \tau_{SR}$ for such $\tau_{SR}$ i.e. $\Delta p$ is determined by the Shockley – Read – Hall recombination and does not depend on the radiative recombination. For these $\tau_{SR}$ values the excitation level $\Delta p$ increases linearly with $\tau_{SR}$ increase and for $\tau_{SR}$ equal to $5 \cdot 10^{-9}$ s and $5 \cdot 10^{-8}$ s $\Delta p$ equals to $2.9 \cdot 10^{11}$ cm$^{-3}$ and $2.9 \cdot 10^{12}$ cm$^{-3}$ respectively. Correlating these values with $q_{pl}$ dependencies shown in Fig. 6, we find that for $b = 0.02$ and $\tau_{SR} = 5 \cdot 10^{-9}$ s the luminescence internal quantum yield $q_{pl}$ in the p-type semiconductor is about 7% and for $\tau_{SR} = 2 \cdot 10^{-8}$ s $q_{pl} = 8\%$. For $b = 50$ the luminescence internal quantum yield $q_{pl}$ rises to 32 % and 37 % respectively. And only for $\tau_{SR} = 10^{-6}$ s and $\Delta p \approx 1.3 \cdot 10^{14}$ cm$^{-3}$ $q_{pl}$ value is close to 93%. Thus, $q_{pl}$ value is close to 100% only at laser excitation levels (obtained in [18]) when the $\Delta p \geq 10^{17}$ cm$^{-3}$. This situation does not correspond to the excitation levels of illuminated SC.

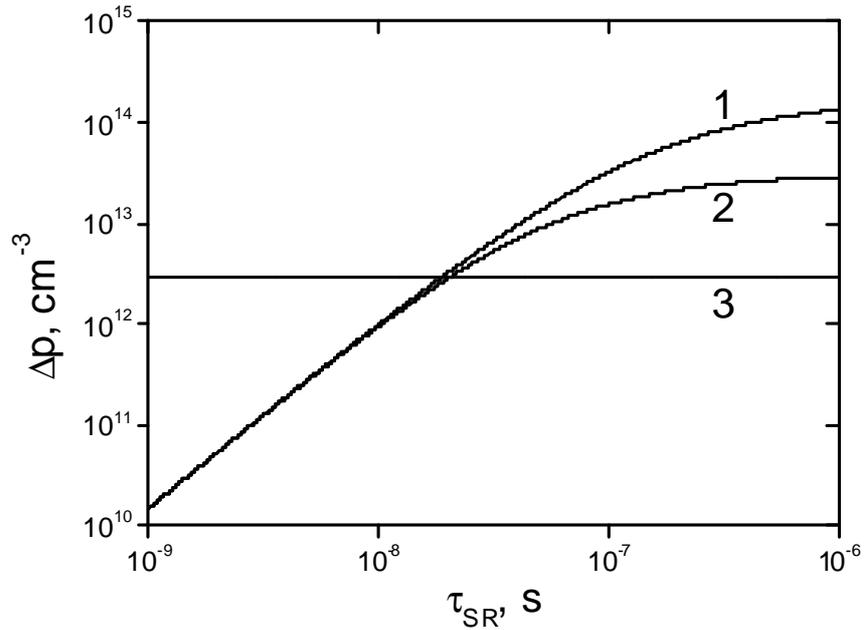

Figure 7. Excitation level versus the Shockley-Read-Hall recombination lifetime. Parameters used: $p_0 = 10^{17}$ cm$^{-3}$, $d = 10^{-4}$ cm, $S = 10^3$ cm/s, $b = 0.02$, $J_{SC} = 30$ mA/cm$^2$, 1 – $A = 10^{-11}$ cm$^3$/s, 2 – $A = 5 \cdot 10^{-10}$ cm$^3$/s, 3 – $A = 5 \cdot 10^{-10}$ cm$^3$/s, $\tau_{SR} = 2 \cdot 10^{-8}$ s.

Note that in the estimates above the absence of the surface recombination influence on the luminescence internal quantum yield was assumed. However, GaAs-AlGaAs system was used [3,14] in the fabrication of structures with high back surface reflectance and heterointerface recombination velocity lies in the range of $10^2 - 10^4$ cm/s [15,16]. For $\tau_b = \tau_r = 10^{-7}$ s, $d = 10^{-4}$ cm, $S_s = 10^3$ cm/s the luminescence internal quantum yield $q_{pl}$ is 33 % (14). For $S_s > 10^3$ cm/s, for example, when $S_s = 10^4$ cm / s we get even less $q_{pl}$ value. In the latter case, even in the absence of bulk Shockley-Read-Hall recombination the luminescence quantum yield is essentially reduced due to the GaAs-AlGaAs heterojunction recombination. Note that GaAs-AlGaAs system is the most matched and has the smallest the lattice constants mismatch. All other heterostructures have greater lattice mismatch and therefore greater $S_s$ values.

For p-n-junction GaAs solar cells the typical $S_0$ and $S_d$ values are of the order of at least $10^5$ cm/s [15]. The luminescence internal quantum yield $q_{pl}$ for such SCs is significantly less than 1, so the effects of light re-emission and reabsorption (photon recycling) can be neglected even for large $\tau_{SR}$ values.

### 2.5. The open circuit voltage dependence on the semiconductor doping level

Using saturation currents [4] and considering the photon recycling the maximum obtainable value of the open circuit voltage in direct-gap semiconductors $V_{OC}$ can be calculated (12). The radiative recombination saturation current density $J_{0r} = 7 \cdot 10^2$ A/cm$^2$ for B-type structures according to [4] was used to calculate the effective radiative recombination parameter $A_{eff}$ from

$$J_{0r} = q A_{eff} d N_c N_v. \qquad (16)$$

For $d = =$ For highly reflective $2 \cdot 10^{-4}$ cm thick type B structures the radiative recombination parameter $A_{eff}$ is about $4 \cdot 10^{-12}$ cm$^3$/s, and for poorly reflecting type A structures (in which, according to [4] $J_{0r} \approx 1 \cdot 10^4$ A/cm$^2$), this parameter is about $10^{-10}$ cm$^3$/s. Substitution of the first value to the formula (12) using the parameters of a semiconductor with a band gap of 1.4 eV (as in [2]) neglecting non-radiative recombination gives a value of 1.142 V for $V_{OC}$, which is only in fourth sign different from the $V_{OC}$ value of 1.145 V [1,2].

The calculated open circuit voltage $V_{OC}$ dependence for type A and type B structures for the case where $\tau_{SR} = 2 \cdot 10^{-8}$ s (curves 1 and 2), $\tau_{SR} = 10^{-6}$ s (curves 3 and 4) and $\tau_{SR} = \infty$ (curves 5 and 6) on the doping level is plotted on the fig. 8. The $A_{eff}$ of $10^{-10}$ cm$^3$/s for type A structures and the $A_{eff}$ of $4 \cdot 10^{-12}$ cm$^3$/s for type B structures were used for figure 8 instead of $A$. The coincidence of $V_{OC}(n_0)$ dependences up to $n_0 = 10^{17}$ cm$^{-3}$ for $\tau_{SR} = 2 \cdot 10^{-8}$ s with the calcu-lated $A_{eff}$ values means that in this doping levels area $V_{OC}(n_0)$ values are determined by non-radiative processes and are independent from the photon recycling. Such dependence appears for $n_0 > 10^{17}$ cm$^{-3}$, but the $V_{OC}(n_0)$ value continues to grow fast enough with $n_0$ increase.

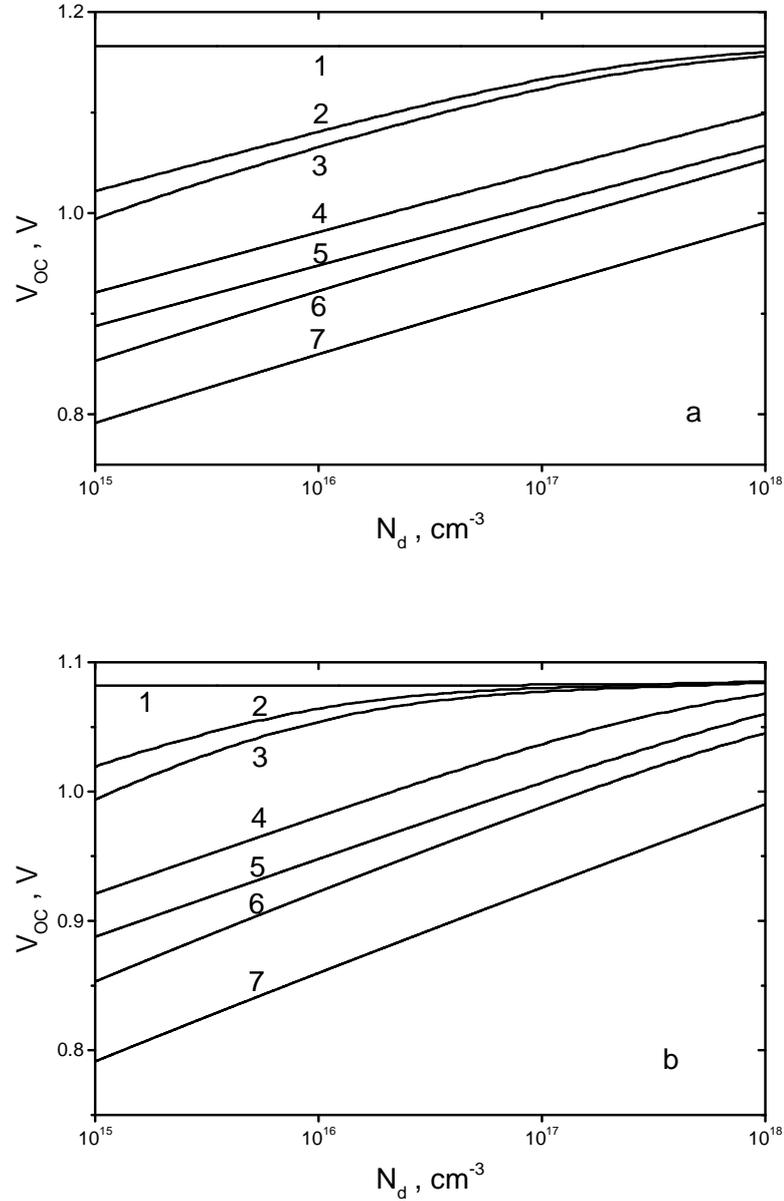

Figure 8. Open-circuit voltage dependence on the doping level. Parameters used: $d = 10^{-4}$ cm, $\varepsilon_s = 12.8$, $S = 1$ cm/s, $T = 300$ K, $E_g = 1.42$ eV, $D = 10$ cm$^2$/s, $S = 1$ cm/s. Figure 8a is calculated for $A_{eff} = 4 \cdot 10^{-12}$ cm$^3$/s, and Figure 8b is calculated for $A_{eff} = 10^{-10}$ cm$^3$/s. Curves parameters for 8: 1 – $\tau_{SR} = 10$ s, $b = 0.02$; 2 – $\tau_{SR} = 10^{-6}$ s, $b = 50$; 3 – $\tau_{SR} = 10^{-6}$ s, $b = 0.02$; 4 – $\tau_{SR} = 2 \cdot 10^{-8}$ s, $b = 50$; 5 – $\tau_{SR} = 2 \cdot 10^{-8}$ s, $b = 0.02$; 6 – $\tau_{SR} = 5 \cdot 10^{-9}$ s, $b = 50$; 7 – $\tau_{SR} = 5 \cdot 10^{-9}$ s, $b = 0.02$.

The $V_{OC}(n_0)$ values differ markedly among themselves for $\tau_{SR}=10^{-6}$ s starting with the doping levels $n_0$ of $10^{16}$ cm$^{-3}$, but $V_{OC}$ value growth with $n_0$ increase continues. The figure shows that $V_{OC}(n_0)$ dependence is growing up to $n_0=10^{18}$ cm$^{-3}$ except of type A structures with $\tau_{SR}=10^{-6}$ s. The $V_{OC}(n_0)$ dependence is stronger than the $V_{OC}(A_{eff})$ dependence of Fig. 9 up to currently maximum Shockley-Read-Hall lifetimes of $10^{-6}$ s. For example, the $V_{OC}(A_{eff})$ value growth for $\tau_{SR}=2\cdot 10^{-8}$ s is equal to 1.5%, while the $V_{OC}(n_0)$ value growth is of 6%. Even for $\tau_{SR}=10^{-6}$ s the $V_{OC}(A_{eff})$ value growth is of 8% and the $V_{OC}(n_0)$ value growth is of 11%, i.e. $V_{OC}$ growth by $n_0$ prevails. It should be noted that nonradiative recombination dominates for $V_{sc} > d/\tau_r$, i.e. even for cases when Shockley-Read-Hall lifetime is not less than radiative lifetime ($\tau_{SR} \geq \tau_r$). For example, for $\tau_{SR}=\tau_r=10^{-7}$ s and $d=1$ μm the $V_{sc}$ value exceeds $d/\tau_r$ value in six times.

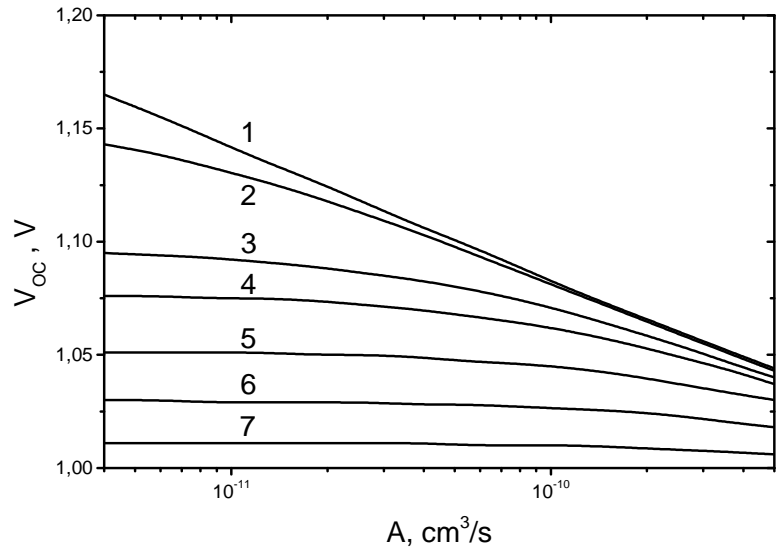

Figure 9. Open circuit voltage $V_{OC}$ dependence on the radiative recombination coefficient A taking into account non-radiative recombination. Parameters used: $n_0=2\cdot 10^{17}$ cm$^{-3}$, $d=2\cdot 10^{-4}$ cm, $T=300$ K, $E_g=1.42$ eV, $D=10$ cm$^2$/s, $S=1$ cm/s, $\varepsilon_s=12.8$, $b=1$.

In conclusion, we note that the use of the $J_{0r}=10^4$ A/cm$^2$ value for poorly reflecting structures and $J_{0r}=7\cdot 10^2$ A/cm$^2$ value for highly reflective structures leads to an overestimation of the photon recycling influence on the open-circuit voltage $V_{OC}$ while radiative and nonradia-tive recombinations make comparable contributions. This is due to the use of 100% luminescence internal quantum yield in the light re-emission and reabsorption (photon recycling) [1,2,4].

If the luminescence internal quantum yield is reduced (per exemplum from 100 to 50%), only 1-2 successive reabsorption and re-emission acts will take place in highly reflective structures [1] instead of 30-40 acts. Thus the $J_{0r}$ value increases reducing the difference between $A$ and $A_{eff}$. As a result re-emission and reabsorption processes influence will be decreased even earlier than the calculations above shows. So these formalisms are correct for $V_{OC}$ calculation in the absence of nonradiative processes only. In intermediate cases of a comparable contribution of the radiative and nonradiative recombination to the open circuit voltage $V_{OC}$, the $V_{OC}$ values are less than values of Fig. 8 and 9.

**2.6. Influence of the electrons and holes capture cross sections on the open-circuit voltage**

Table 1
The electrons and holes capture by the deep recombinational level cross sections ratio $b$ influence on the open-circuit voltage $V_{OC}$ for different Shockley-Read-Hall lifetimes $\tau_{SR}$.

| $b$ | $\tau_{SR}$, c | $V_{OC}$, B |
|---|---|---|
| 0.02 | $5\ 10^{-9}$ | 0.945 |
| 0.02 | $8\ 10^{-8}$ | 1.007 |
| 50 | $5\ 10^{-9}$ | 1.024 |
| 50 | $2\ 10^{-8}$ | 1.051 |
| 0.02 | $10^{-6}$ | 1.080 |
| 50 | $10^{-6}$ | 1.082 |

Since the recombination velocity in the SCR $V_{SC}$ and, therefore, the saturation current density depend on the deep recombination level charge carrier capture cross sections ratio $b$, the open-circuit voltage $V_{OC}$ will also depend on this ratio. According to (10) and (12), for $b>1$ $V_{SC}$ is decreased (so $V_{OC}$ is increased) and for $b<1$ $V_{SC}$ is increased (so $V_{OC}$ is decreased) in n-type semiconductors. The situation is reversed in p-type semiconductor, i.e. when $b>1$ $V_{OC}$ is decreased and when $b<1$ $V_{OC}$ is increased. Table 1 shows the n- and p-type gallium arsenide $V_{OC}$ values, calculated for various $b$ values. The table shows that $V_{OC}$ values can be significantly adjusted by the $b$ value. Particularly, the $V_{OC}$ value in n-type GaAs for $b=0.02$ is smaller than the value in the p-type even when the Shockley-Read-Hall lifetime $\tau_{SR}$ in the n-type exceed $\tau_{SR}$ value in the p-type. Thus, the $b$ value should be considered in selection of the material for highly efficient solar cells.

**2.7. Influence of the doping type and level on the short-circuit current**

An analysis of the doping type and level influence on the short-circuit current magnitude is carried out on the typical direct-gap semiconductor - gallium arsenide. Short-circuit current density under AM1.5 conditions in the case of full absorption of the incident light, neglecting SC illuminated surface shading by electrodes is defined by the relationship:

$$J_{SC} = (1 - R_s)(1 - m) \int_{E_g}^{\infty} J_{AM1.5}(E_{ph}) q_{SC}(\alpha) dE_{ph} . \quad (17)$$

Here $R_s$ is the light reflection coefficient on the SC illuminated surface, $m$ is the illuminated surface shading by the contact grid factor, $J_{AM1.5}$ is the spectral photocurrent density for AM1.5 illumination.

Table 2.
The radiative recombination parameter $A$ and short-circuit current density $J_{SC}$ as a function of acceptors concentration $N_a$ (p-type) and donors concentration $N_d$ (n-type) in GaAs.

| $N_a$, см$^{-3}$ | $A$, см$^3$/с | $J_{SC}$, мА/см$^2$ |
|---|---|---|
| 1.6 10$^{16}$ | 6.65 10$^{-10}$ | 31.663 |
| 2.2 10$^{17}$ | 7.32 10$^{-10}$ | 31.376 |
| 4.9 10$^{17}$ | 11.9 10$^{-10}$ | 30.642 |
| 1.2 10$^{18}$ | 12.5 10$^{-10}$ | 29.308 |
| 2.4 10$^{18}$ | 14.0 10$^{-10}$ | 27.271 |

| $N_d$, см$^{-3}$ | $A$, см$^3$/с | $J_{SC}$, мА/см$^2$ |
|---|---|---|
| 5 10$^{13}$ | 9.84 10$^{-10}$ | 31.813 |
| 5.9 10$^{17}$ | 5.69 10$^{-10}$ | 27.400 |
| 2 10$^{18}$ | 3.51 10$^{-10}$ | 24.937 |
| 3.3 10$^{18}$ | 1.82 10$^{-10}$ | 24.815 |

Since the gallium arsenide light absorption coefficient depends essentially on both doping level and conductivity polarity, this dependence should be taken into account in the sort-circuit current $J_{SC}$ calculation. The results of this calculation for the gallium arsenide are shown in Ta-ble 2. Table 2 shows that $J_{SC}$ value for the p-type semiconductor increases with the doping level increase. This growth is associated with an increase in fundamental absorption edge blur with the increase in doping level (see Fig. 1). At the same time, the $J_{SC}$ value for n-type semiconductor decreases with the increase in doping level due to the conduction band filling by carriers (Burstein-Moss effect).

## 3. Direct-gap semiconductors photoconversion efficiency

Photoconversion efficiency calculation was fulfilled for GaAs SC as an example. The equation for I-V curve was written as

$$J(V) = J_{SC} - q\left(\frac{d}{\tau_b} + S_s\right)\frac{n_i^2}{n_0}\left[\exp\left(\frac{qV}{kT}\right) - 1\right] - q\frac{\kappa L_D b^{-1/2} n_i}{\tau_{SR}\sqrt{\frac{1}{2}\ln\left(\frac{qn_0(d/\tau_b + S_s)}{J_g b}\right)}}\left[\exp\left(\frac{qV}{2kT}\right) - 1\right]. \quad (18)$$

The photovoltage at maximum takeoff power $V_m$ can be found from $d\{VJ(V)\}/dV = 0$ condition. The photocurrent density $J_m$ for this case can be calculated by the substitution of $V_m$ into (18). Finally photoconversion efficiency can be written as

$$\eta = \frac{J_m V_m}{P_S}, \qquad (19)$$

where $P_S$ is the illumination power density per unit area.

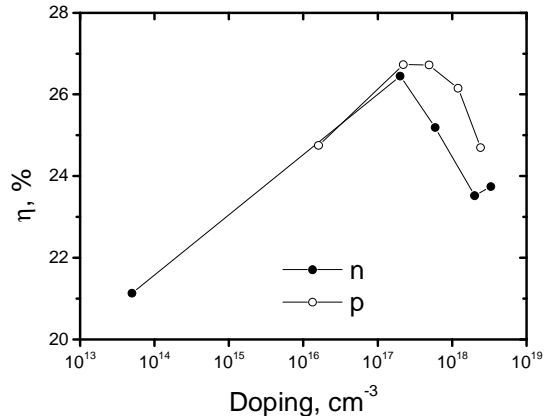

Figure 10. The photoconversion efficiency dependence on the doping level. 1 for p-type and 2 for $n$-type.

Fig. 10 shows the calculated using (19) GaAs photoconversion efficiency values versus the base material doping level for both conductivity polarities and the following Shockley-Read-Hall lifetimes $\tau_{SR}$: $5 \cdot 10^{-9}$ s for p-type and $2 \cdot 10^{-8}$ s for the n-type. Zero $R_s$ and $m$ values were applied.

One can see from the figure that photoconversion efficiency $\eta$ increases with the doping level growth for both p-type and n-type. This increase is first of all due to an increase of $V_{OC}$. At larger doping levels $\eta$ decreases due to the interband Auger recombination predominance. Lower n-type SC $\eta$ values in this region are associated with lower $J_{SC}$ values. Maximum $\eta$ values for both p- and for n-type conductivity are realized at doping levels of about $10^{17}$ cm$^{-3}$ and are close to the value of about 27% for typical $\tau_{SR}$ values. Thus this doping level is the optimal for the maximum efficiency $\eta$ achievement in gallium arsenide.

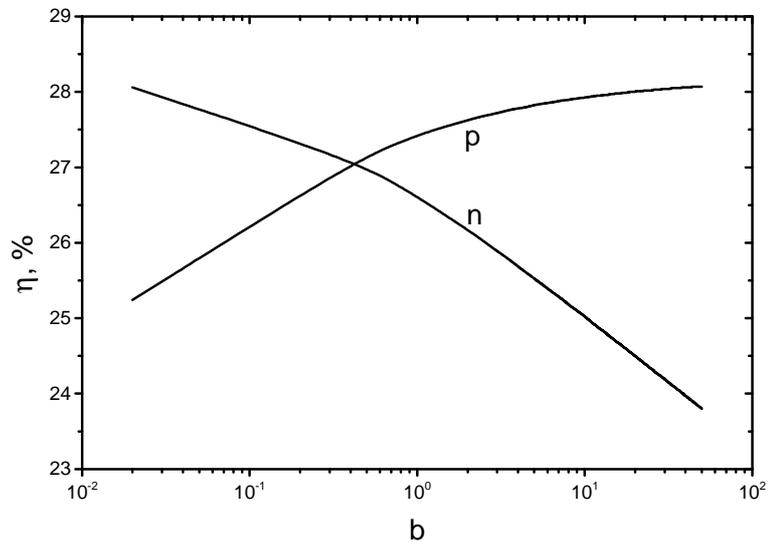

Figure 11. The photoconversion efficiency versus the electrons and holes capture cross sections ratio $b$ in GaAs. Parameters used: $n_0(p_0) = 2 \cdot 10^{17}$ cm$^{-3}$, $A = 5 \cdot 10^{-10}$ cm$^3$/s, $E_g = 1.42$ eV, $d = 10^{-4}$ cm, $\varepsilon_s = 12.8$, $S = 10^3$ cm/s; p-type: $J_{SC} = 31.6$ mA/cm$^2$, $\tau_{SR} = 5 \cdot 10^{-9}$ s, $D = 50$ cm$^2$/s; n-type: $J_{SC} = 30.9$ mA/cm$^2$, $\tau_{SR} = 2 \cdot 10^{-8}$ s, $D = 10$ cm$^2$/s.

Fig. 11 shows calculated $\eta$ values versus the $b$ value for p- and n-type gallium arsenide SCs. It can be seen that the maximum $\eta$ values for p- and n-type gallium arsenide SCs are very similar, but the minimum values are much smaller for n-type. It is here evident that n-type Shockley-Read-Hall lifetime is less then p-type lifetime. It leads to a significant $V_m$ value decrease.

Therefore, to achieve both high open-circuit voltage value and high short circuit current density, i.e. to get the maximum photoconversion efficiency $\eta$ we must use p-type semiconduc-tor as the base material. Note that diffusion length of electrons in the p-type base material is much greater than the diffusion length of holes in the n-type base material. It is possible that the p-type GaAs disadvantage compared to the n-type is the smaller Shockley-Read-Hall lifetime $\tau_{SR}$. However, according to [7, 8], typical p-type lifetimes $\tau_{SR}$ are not so small compared to the n-type, so it is prematurely to do the final conclusions about the n-type.

We also note that even for the maximum Shockley-Read-Hall lifetime of $10^{-6}$ s achieved the photon recycling neglect leads to a not more than 8% reduction in the case of highly reflective structures, and to a not more than 3% reduction in the case of poorly reflecting structures. These differences are estimated to be not so great compared to, for example, the summarized influence of the surface recombination velocity and electrons and holes capture cross sections by the recombination level ratio, especially with the doping level influence on the $V_{OC}$ value. Therefore, the influence of these factors on the photoconversion efficiency must be considered in the first place.

## 4. Photoconversion efficiency analysis features for other semiconductors (the indirect and direct band)

The relationships above can be used to optimize the other direct-gap semiconductors solar cells (including A3B5 group) parameters. In the first place, the article includes the general relations allowing the fulfilling of a specified analysis. Secondly, a number of these semiconductors parameters are very close to the GaAs parameters. In particular, the electrons and holes effective masses that determine the mobility and diffusion coefficients in these semiconductors are very close to the GaAs values. Consequently, the densities of states in the conduction band and the valence band, as well as radiative recombination coefficients are close. The Shockley-Read-Hall lifetimes should be close as well. The single difference is the relation between the radiative and nonradiative processes in the semiconductor volume for these materials that changes in favor of the radiative processes dominance with the bandgap $E_g$ decrease. So the photon recycling growth for narrow-gap semiconductors is possible. However, the following two conditions have to be performed. First, this situation is possible only for the small surface recombination velocity values. But the fulfillment of this condition is very difficult. The AlGaAs - GaAs systems, as mentioned, have the interfaces of very high quality, with the lowest recombination velocities. Heterojunction interfaces of other semiconductors are not perfect and the recombination velocities at the boundaries are much higher than for the AlGaAs - GaAs system. This fact eliminates the role the radiative processes.

Second, it is difficult to realize good light reflection conditions in the each cell of a multijunction SC because it needs a certain ratio between the refractive indices of the neighbor cells. This condition limits the photon recycling again. Finally, the open circuit voltage growth for the photon recycling is inessential for the actual Shockley-Read-Hall recombination times. Regarding to the faster open-circuit voltage decrease than bandgap decrease with bandgap narrowing, the error due to the neglect of photon recycling in the total open circuit voltage value for the typical $\tau_{SR}$ values will be small.

### 4.1. Indirect semiconductors (silicon)

The relations obtained can be used in a number of cases to optimize the indirect semiconductors (for example, silicon, germanium) solar cells parameters, in which the radiative recombination probability is small. The silicon and germanium are particularly used as a third element in some three-junction SCs. This is especially important for monocrystalline and multicrystalline silicon solar cells as silicon solar modules and batteries are widely used. Moreover, monocrystalline and multicrystalline silicon SCs produce the prevalent part of electricity generated by direct solar energy conversion for now.

Until recently, high-efficiency silicon solar cell parameters were modeled using the numerical solution of the drift and diffusion equations [9,12]. There were also attempts [4,17] to model silicon solar cells efficiency in the same approximations as gallium arsenide SCs, i.e. taking into account the photon recycling (re-emission and absorption of photons) in highly reflective structures or structures with multiple reflection (equal to the raise of the light absorption optical thickness). The photon recycling for silicon is assumed to be much lower than for gallium arsenide, where the luminescence internal quantum yield is assumed to be near 100% (as demonstrated for the high excitation case [3]). For example, the maximum photoluminescence external quantum yield of just 6.1% was obtained [18] at room temperature. To estimate the luminescence internal quantum yield $q_{pl}$ for silicon, we use the equation (14) with the next interband Auger recombination parameter [11,19]:

$$R_{Auger} = C_p (p_0 + \Delta p)^2 + C_n (p_0 + \Delta p)\Delta p, \quad (20)$$

where $C_p = 10^{-31}$ cm$^6$/s and $C_n = \left(2.8 \cdot 10^{-31} + \dfrac{2.5 \cdot 10^{-22}}{\Delta p^{0.5}}\right)$ cm$^6$/s. The second term in the expression for $C_n$ includes the spatial correlation due to Coulomb interaction effect in the distribution of two electrons and one hole needed to act Auger recombination.

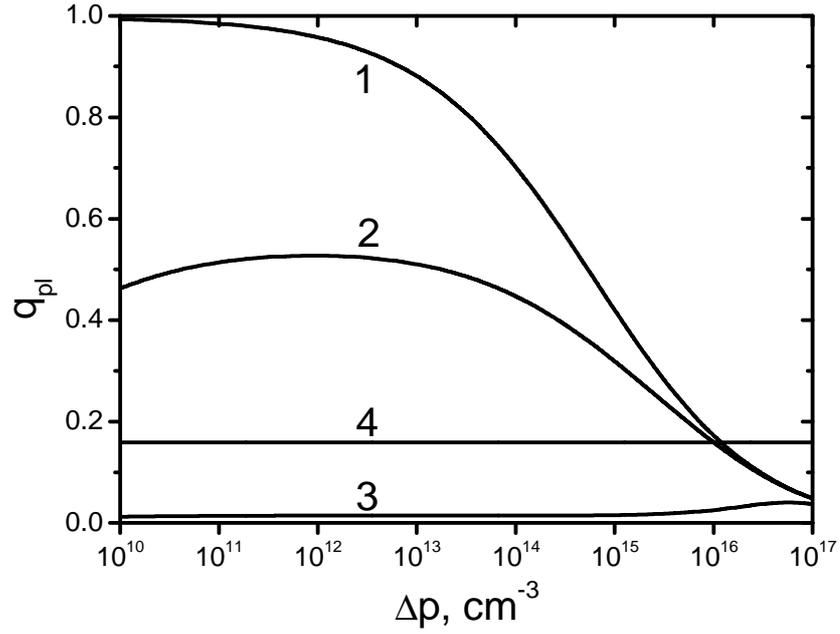

Fig. 12. The luminescence internal quantum yield dependence on the excitation level in silicon. Parameters used: $d$ =200 μm, T=298 K; 1 – $p_0$=3·$10^{16}$ cm$^{-3}$, $\tau_{SR}$=$10^2$ s, $S$=$10^{-3}$ cm/s; 2 – $p_0$=8·$10^{16}$ cm$^{-3}$, $\tau_{SR}$=3.8·$10^{-2}$ s, $S$=10 cm/s; 3 – $p_0$=1·$10^{16}$ cm$^{-3}$, $\tau_{SR}$=$10^{-3}$ s, $S$=60 cm/s.

The $q_{pl}(\Delta p)$ calculation results for the case of silicon are shown in Fig. 12. The value of 6 • 10-15 cm$^3$/s [20] was used as the radiative recombination parameter at T = 300 K. It can be seen from the figure (see curve 1), that the highest value of 99.3% is realized at $\Delta p$ = $10^{10}$ cm$^{-3}$ and $p_0$ = $10^{14}$ cm$^{-3}$ for the case of interband Auger recombination as the only nonradiative recombination channel. In the case of minimal recombination parameters ($\tau_{SR}$ = 38 ms $S$ = 0.25 cm/s) [21,22] achieved in the silicon, the largest $q_{pl}$ value is 52.7% (curve 2). For the silicon solar cells parameters, providing a record efficiency equal to 25% [23], the highest $q_{pl}$ value is 4% (curve 3). At $\Delta p \geq 10^{16}$ cm$^{-3}$, there is a strong $q_{pl}$ drop associated with the increased nonradiative Auger recombination contribution. Since a typical $\Delta p$ value for a silicon solar cell under illumination is greater then $10^{16}$ cm$^{-3}$, as show the calculation (15) $q_{pl}$ value for this situation, as Figure 12 shows, does not exceed 25%. It means that the photon recycling can be neglected in the silicon solar cell parameters (including maximum achievable) simulation.

Note, that the formulae above are not suitable for the silicon solar cells efficiency calculation in the case of over 20% efficiency. The matter is that these solar cells are manufactured from high quality monocrystalline silicon with the Shockley-Read-Hall lifetime not less than 1 ms. Therefore, the diffusion length $L$ for such SCs can reach more than one millimeter, and this length substantially exceeds the base thickness $d$. The electron-hole pairs excess concentration in the open circuit mode, as has been said above, reaches values of the order of $10^{16}$ cm$^{-3}$. If the base doping is less than this value, the open circuit voltage $V_{OC}$ in the p-type semiconductor is determined by the following expression [11]

$$V_{OC} \cong \frac{kT}{q}\ln\left(\frac{\Delta p}{n_0}\right) + \frac{kT}{q}\ln\left(1 + \frac{\Delta p}{p_0}\right). \quad (21)$$

The greater $\Delta p / p_0$ ratio is in comparison to unity, the greater is the open circuit voltage in a highly efficient silicon solar cell as compared to the case where $L < d$. This is associated not only with a larger $\Delta p$ value, but also with the noticeable or even comparable to the illuminated surface voltage value contribution of the back surface to the open-circuit $V_{OC}$. Such a situation is never realized in direct-gap semiconductors, such as gallium arsenide, due to quite a small nonequilibrium charge carriers lifetime and $\Delta p \ll p_0$.

The 25% record photoconversion efficiency in silicon under AM1.5 conditions is set [23] especially for the case above where formula (21) is correct. In addition to an almost complete incident light absorption by the SC implementation (that can be achieved through the use of geometric relief), the main problem is to minimize the surface recombination velocity values on the illuminated and back surfaces. The illuminated surface recombination velocity was minimized by the thermal oxide with a low density of surface states growth and the back surface recombination velocity minimization was achieved by the isotype junction creation by Green [23]. There are recent researches solving this problem by $\alpha - Si:H - Si$ heterojunctions usage [24,25]. In this case the dangling bonds passivation by the hydrogen for the surface recombination velocity $S$ minimization is necessary. The highest 24.7% efficiency of such solar cells was obtained in [25]. It is necessary to determine the $\Delta p$ value from the equation (15) and substitute this $\Delta p$ in (21) to find the $V_{OC}$ value in this case.

Fig. 13 shows the $V_{OC}$ dependence on the base doping level. Curve 1 corresponds to the limiting parameters, when the Shockley-Read-Hall recombination velocity and the surface recombination velocity are negligible little compared to the interband Auger recombination velocity. Curve 2 corresponds to the case of actual silicon parameters ($\tau_{SR} = 38$ ms, $S = 10$ cm/s), and curve 3 simulates the $V_{OC}$ value for the silicon solar cell with a record efficiency of 25% [23]. All curves have the maxima which lie at $p_0 \leq 10^{17}$ cm$^{-3}$ as can be seen from the figure. $V_{OC}$ growth is associated with a relative decrease in the second term (21) contribution with $p_0$ increase in the region of $p_0 < 10^{17}$ cm$^{-3}$. $V_{OC}$ decrease is associated with $R_{Auger}$ growth for $p_0 \geq 10^{17}$ cm$^{-3}$. The calculated by (21) open circuit voltage value can exceed 0.740 V for the record SC parameters. Conventional silicon solar cells with $L < d$ can have such a $V_{OC}$ value only under concentrated illumination.

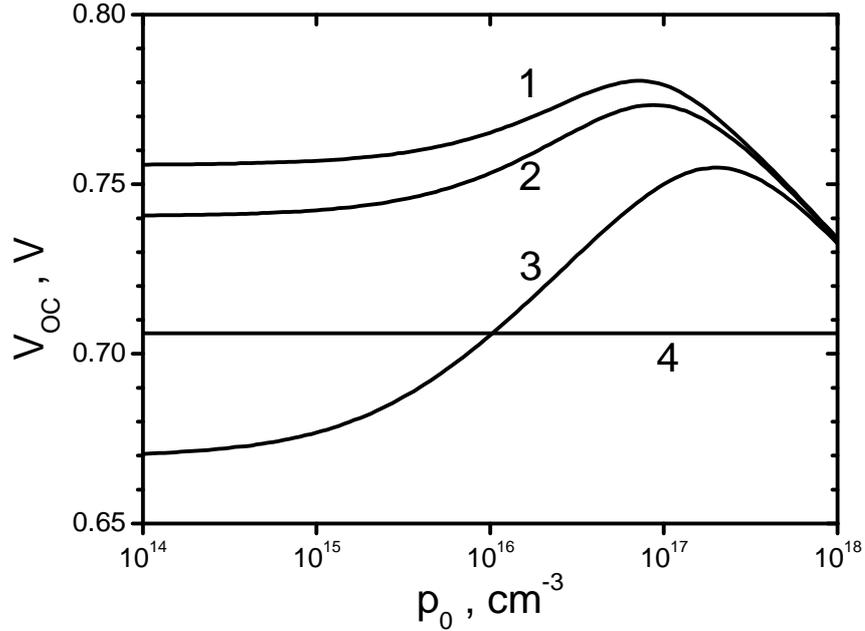

Figure 13. Open circuit voltage dependence on the doping level. Parameters used $d = 200$ µm, $T = 298$ K. The same $\tau_{SR}$ and $S$ values as for Figure 12. Short-circuit current density $J_{SC}$ was assumed: 1 – 43.5 mA/cm², 2 – 42.7 mA/cm², 3 – 42.7 mA/cm².

Equation (21) is a quadratic equation with regarding $\Delta p$ and its solution has the following form:

$$\Delta p = -\frac{p_0}{2} + \sqrt{\frac{p_0^2}{4} + \exp\left(\frac{qV_{OC}}{kT}\right)}. \qquad (22)$$

Replacing the $V_{OC}$ in (22) by the applied forward bias $V$, we get the $\Delta p$ to $V$ relation. It allows to write current-voltage characteristic on the basis of expression (15)

$$J(V) = J_{SC} - J_{rec}(V). \qquad (23)$$

Further, we find $V_m$ from the maximum takeoff power condition $d(VJ(V))/dV = 0$ and its substitution in (23) allows the $J_m$ value determining. As a result, we obtain the silicon solar cell photoconversion efficiency under AM 1.5 illumination:

$$\eta = \frac{J_m V_m}{P_S}. \qquad (24).$$

Fig. 14 shows the photoconversion efficiency dependence on the Shockley-Read-Hall lifetime $\tau_{SR}$. The basic parameters are the same as previous figure parameters. As can be seen from the figure, for sufficiently large $\tau_{SR}$ values $\eta$ reachs maximum and saturates. The saturation comes earlier for the larger the $S$ values.

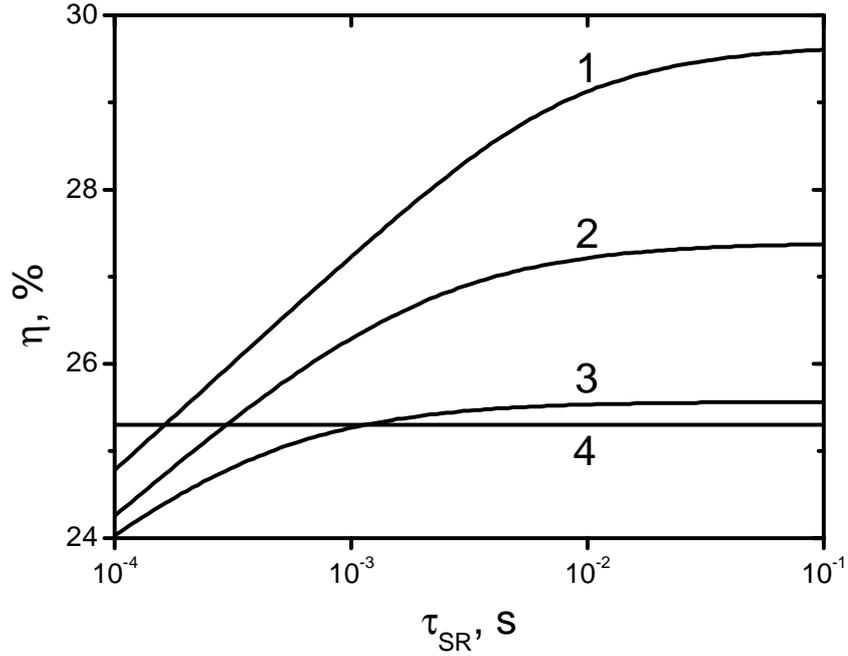

Figure 14. Photoconversion efficiency $\eta$ versus lifetime $\tau_{SR}$. Parameters used: $d = 200$ µm, T = 298 K. The same $S$ values as for Figure 13. Doping level was assumed to be $10^{16}$ cm$^{-3}$.

Fig. 15 shows the $\eta$ dependence on the base doping level $p_0$, which is calculated with the same basic parameters. It can be seen that all three curves have maxima. The lower $p_0$ values in maxima are realized for smaller Shockley-Read-Hall recombination and surface recombination contributions in complete recombination. Maximum attainable $\eta$ value for such parameters is 29.2%. If the solar cell thickness decreases from 200 to 50 µm, the limiting $\eta$ value increases to 30.4%.

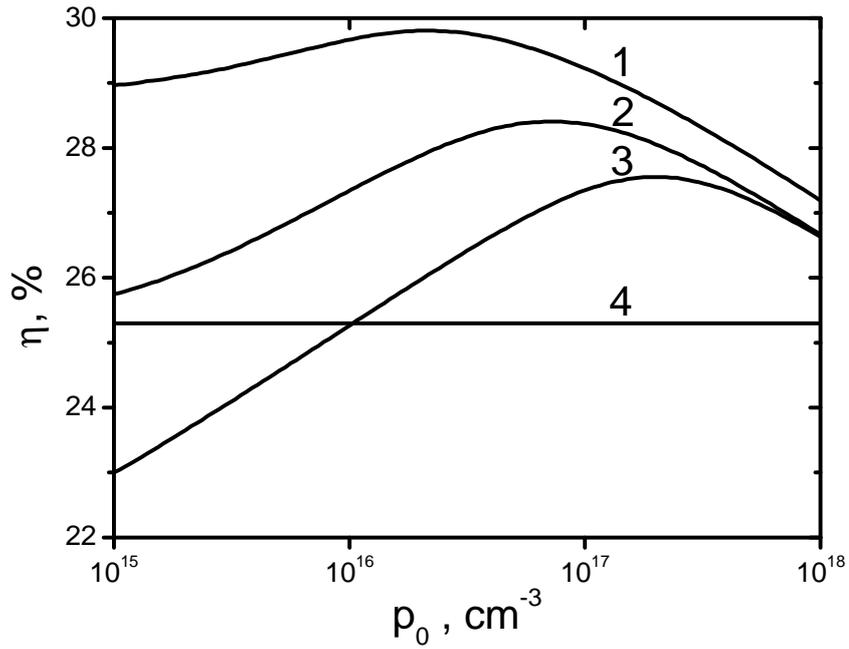

Fig. 15. Photoconversion efficiency $\eta$ versus the doping level. Parameters used: $d = 200$ µm, $T = 298$ K. The same parameters as for Figure 13.

Table 3.

Experimental and calculated values of the main silicon solar cells parameters for three different samples [23-25]

| Number cell | $J_{SC}^{exp}$, mA/cm² | $V_{OC}^{exp}$, V | $V_{OC}^{th}$, V | $\eta_{exp}$, % | $\eta_{th}$, % |
|---|---|---|---|---|---|
| 1 | 427 | 0.706 | 0.706 | 25 | 25.3 |
| 2 | 394 | 0.745 | 0.745 | 23.7 | 23.8 |
| 3 | 395 | 0.750 | 0.750 | 24.7 | 24.9 |

Next, let us to write theoretical equations for open-circuit voltage and photoconversion efficiency versus base doping level for 25%-effective record silicon solar cell [23], as well as for silicon solar cells with surface recombination, minimized by the $\alpha - Si:H - Si$ heterojunctions [24, 25] using the formalism above. First the recombination parameters for $V_{OC}$ of 706 mV, obtained in [23] are calculated. They are determined from the curve 3 in Fig. 13 intersection with the experimental value of 706 mV. Next we substitute the parameters, for which this coincidence was obtained, and experimental short-circuit current density value equal to 42.7 mA/cm², in the expression for $\eta$. In result $\eta$=25.3% is achieved. The relevant parameters for the $\alpha - Si:H - Si$ heterojunction solar cells [24, 25] were calculated similarly. The final results are presented in

Table 3. It can be seen from the table, that the calculated open-circuit voltage values coincide with the experimentally obtained in all cases. The matching error between the calculated and the experimental $\eta$ values is less than 1%. These results demonstrate the adequacy of the theoretical model proposed in this paper relative to the experimental results obtained in [23-25].

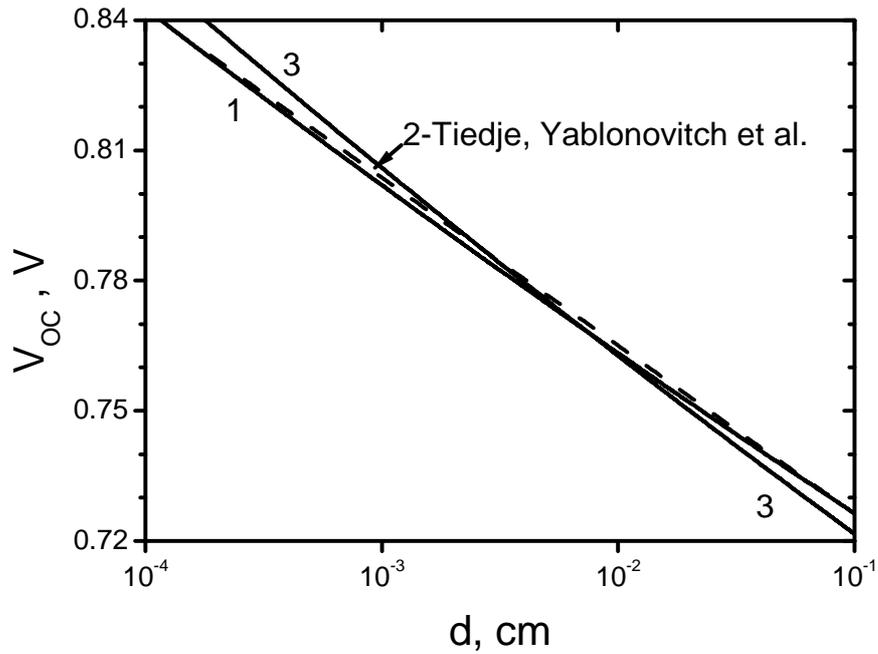

Figure. 16. Calculated open circuit voltage versus the thickness of the limit silicon solar cell obtained in [17] (curve 1), calculated using the parameters of [17] (curve 2) and calculated using the parameters of our papers [19,20] (curve 3).

In conclusion, we compare the results obtained in the present article with the results of the research [17]. In this paper, Fig. 8 shows the results of the open circuit voltage calculation depending on the silicon solar cell thickness taking into account radiative recombination and interband Auger recombination. These dependences are calculated using expressions (21) and (15). Firstly, let us correct the parameters to the parameters of [17]. Thus, we assume that the intrinsic charge carriers concentration in silicon (for T = 300 K) equals to $1.45 \cdot 10^{10}$ cm$^{-3}$, and the radiative recombination coefficient equals to $2.5 \cdot 10^{-15}$ cm$^3$/s [17]. Let us also omit the $2.5 \cdot 10^{-22}/\Delta p^{0.5}$ term in $C_n$ which is absent in [17]. The comparison of the $V_{OC}(d)$ dependencies calculated using formulae (15) and (21) of our work (Figure 16, curve 1) with those achieved in [17] (Figure 8) (see Fig. 16 of present paper, curve 2) show the good agreement.

This figure also shows the $V_{OC}(d)$ curves constructed using formulae (15) and (21) of the present paper and the parameters of silicon from [19, 20] (curve 3). There is a good comparison of the curves (1) and (3), that practically coincide for the actual silicon solar cell thickness (∞100 μm), although they are different for small and large SC thicknesses. A little higher $V_{OC}(d)$ values at small thickness values are associated with the use of lower holes and electrons concentration in silicon equal to 8.5 • $10^9$ cm$^{-3}$. Lower $V_{OC}(d)$ values at large thickness values are explained by the greater radiative recombination parameter at room temperature (6 • $10^{-15}$ cm$^3$/s) and by taking into account the additional term in the relation for $C_n$.

The limiting silicon solar cells photoconversion efficiency of our article correlates with that obtained in [17], and is consistent with the conclusions of [26]. However, the photon recycling for calculating the silicon solar cells parameters (including the maximum possible and record) can be neglected, as shown by the results of our analysis, in contrast to [4, 17]. As a result, the calculation is greatly simplified and can be performed in the traditional approximations.

## 5 Conclusions

An approach to optimize the SC based on direct and indirect-gap semiconductors in order to obtain the maximal photoconversion efficiency is proposed.

Analysis shows the secondary role of photon recycling even in highly reflective structures for mass production GaAs parameters, particularly typical Shockley-Read-Hall recombination lifetime $\tau_{SR}$ values. The open circuit voltage $V_{OC}$ enhancement by the doping level optimization is shown to be much more critical for photoconversion efficiency growth. The optimal doping level of about $10^{17}$ cm$^{-3}$ is determined. The maximal photoconversion efficiency $\eta_{max}$ for such condition is near 27%.

The doping level and polarity type influence on the GaAs absorption coefficient and photoconversion efficiency is studied. It is shown that the conduction band filling by electrons with increase in the donor doping level (Burstein-Moss effect for n-type semiconductor) leads to a decrease in the short-circuit current. The fundamental band edge blur increase (Urbach shift) leads to an increase in the short-circuit current for p-type semiconductor.

The influence of the electron and hole capture by the deep recombination level cross section ratio influence on the open-circuit voltage and, consequently, the photoconversion efficiency was demonstrated.

Obtained formalism allows to analyze and to optimize SC parameters for other direct-gap semiconductors, particularly A3B5 semiconductors.

The open circuit voltage and photoconversion efficiency generation features for monocrystalline silicon solar cells with more than 1 ms Shockley-Read-Hall lifetime were considered.

A formalism describing quantitatively the experimental results for high-efficiency silicon solar cells using various surface recombination velocity minimization techniques was proposed.

The approach of this research allows to predict the expected solar cells (for both direct-gap and indirect-band semiconductor) characteristics if material parameters are known.